\documentclass[aps,superscriptadress,preprint]{revtex4}
\usepackage{graphicx}
\usepackage{amsmath}
\usepackage{mathrsfs}
\usepackage{amssymb}
\usepackage{booktabs}
\usepackage{tabularx}
\usepackage{textcomp}
\usepackage{feynmf}
\usepackage{color}
\begin{document}
\title{Baryon Number Fluctuations in Two Flavor NJL Model}

\author{Ameng Zhao$^{1}$}~\email[]{Email:zhaoameng@cxxy.seu.edu.cn}
\author{Hongshi Zong$^{3,4,5}$}
\address{$^{1}$ Department of Foundation, Southeast University Chengxian College, Nanjing 210088, China}
\address{$^{3}$ Department of Physics, Nanjing University, Nanjing 210093, China}
\address{$^{4}$ Joint Center for Particle, Nuclear Physics and Cosmology, Nanjing 210093, China}
\address{$^{5}$ State Key Laboratory of Theoretical Physics, Institute of Theoretical Physics, CAS, Beijing 100190, China}

\begin{abstract}
Baryon number fluctuations are believed to be good signatures of the QCD phase transition and its CP. Since the fluctuations are proportional to the various order baryon-number susceptibilities and the quark-number density is related to the dressed quark propagator, then by the two flavor NJL model we can calculate the moments of the up and down quarks. By comparing the two flavor NJL model results with the experimental data of $S\sigma$ and $\kappa\sigma^2$ for the net-proton distributions of Au+Au collisions by STAR, we found that the NJL results fit the experiments better at large collision energies and show an obvious fluctuation at small energies. And this fluctuation reflect that the two flavor NJL model is sensitive to the parameters of  temperature $T$ and quark chemical potential $\mu$ at small collision energies.

%\\\newline


\bigskip

\end{abstract}

\maketitle

\section{Introduction}
To search for the CP and phase boundary in the QCD phase diagram, RHIC has undertaken its first phase BES(Beam Energy Scan) Program\cite{b2,b2a,b2b,b2c,b2d}. Since the moments of the distributions of conserved quantities, for example net-baryon number, in the relativistic heavy ion collisions are related to the correlation length $\xi$ of the system\cite{b3,b3a}, they are believed to be good signatures of the QCD phase transition and its CP.

It is known that the moments of the baryon number are proportional to the various order baryon-number susceptibilities, and their relations are shown in \cite{b4}. It can be seen that when relating the susceptibilities to the moments a volume term appears. In order to cancel the volume term, the products of the moments, $S\sigma$ and $\kappa\sigma^2$, are constructed as the experimental observables. The results in RHIC of these observables show a centrality and energy dependence\cite{b5}. When studying the quark numbers at finite temperature and chemical potential, it is found that the quark-number density is determined by the corresponding dressed quark propagator only\cite{b6,b6a}. Then by calculating the derivatives of the quark-number density with respect to $\mu$, we can calculate the quark-number susceptibilities at finite temperature and chemical potential.

In this paper, we obtain the dressed quark propagator in the Nambu-Jona-Lasinio(NJL) model. The NJL model\cite{b7,b7a,b7b}, as a widely adopted phenomenological model of QCD, is used to study the dynamical chiral symmetry breaking(DCSB) and the interaction between hadrons. The NJL model keeps the basic symmetries of QCD and simplifies the interactions in QCD to four-body interactions.
\section{Moments by NJL Model}
\label{two}

The commonly used Largrangian of the two-flavor NJL model is
\begin{equation}
\mathcal{L}=\bar{\psi}(i\gamma^{\mu}\partial_{\mu}-m)\psi+G[(\bar{\psi}\psi)^2+(\bar{\psi}i\gamma_5\vec{\tau}\psi)^2],
\label{1}
\end{equation}
where $m=m_u=m_d$, $\psi=(\psi_u,\psi_d)^T$ and $G$ is the effective coupling strength of the four-point quark interaction. With the mean field approximation of Eq.{\ref{1}}, the effective quark mass $M$ is
\begin{equation}
M=m-2G<\bar{\psi}\psi>,
\label{2}
\end{equation}
where the quark condensate at finite temperature($T$) and chemical potential($\mu$) is defined as
\begin{eqnarray}
\begin{split}
<\bar{\psi}\psi>=&-TN_cN_f\sum^{+\infty}_{-\infty}\int^{\Lambda}\frac{d^3p}{(2\pi)^3}\frac{4M}{p^2+M^2+\widetilde{\omega}^2_n}\\
=&-\frac{N_cN_fM}{2\pi^2}\int^{\Lambda}dp\frac{p^2}{\sqrt{p^2+M^2}}(th(\frac{\sqrt{p^2+M^2}+\mu}{2T})+th(\frac{\sqrt{p^2+M^2}-\mu}{2T})),
\end{split}
\end{eqnarray}
where $N_c=3$ is the number of color and $N_f=2$ is the number of flavor, $\widetilde{\omega}_n=2\pi nT+i\mu$ and the NJL model three-momentum non-covariant cut-off is $\Lambda$. In this paper, we adopt a varying coupling strength $G$\cite{b8,b8a,b8b,b8c,b9}, which is used to reproduce the Lattice results at finite temperature\cite{b9}:
\begin{equation}
G=G_1+G_2<\bar{\psi}\psi>,
\label{g2}
\end{equation}
where $G$  in the NJL model represents the effective gluon propagator, $G_2<\bar{\psi}\psi>$ reflects the contribution of the two-quark condensate to the gluon propagator and $G_1$ reflects all the other condensate contributions. Since the quark propagator and gluon propagator are coupled with each other by QCD and the quark propagator in Nambu and Wigner phase are very different, then the corresponding gluon propagators in these two phases should be different too. At the same time Lattice results have shown that the gluon propagator, rather than a constant, evolves with temperature. That is to say a constant $G$ does not meet these requirements. In \cite{b8c} they investigate how to extract the feedback of quark from gluon propagator and get Eq.{\ref{g2}}.The parameter set used in this paper is $m=5.6 MeV$, $\Lambda=587.9 MeV$, $G_1=5.564\times10^{-6} MeV^{-2}$ and $G_2=-3.16\times10^{-14} MeV^{-5}$, which is proved to be successful in fitting the quark condensate to the Lattice results at finite temperature\cite{b9}. Then by Eq.{\ref{2}} we could get the effective quark mass $M$ as a function of $T$ and $\mu$. Since the quark number density of $N_c$ color and $N_f$ flavor is\cite{b6,b6a}
\begin{eqnarray}
\begin{split}
n_q=&<\psi^+\psi>=-TN_cN_f\sum^{+\infty}_{-\infty}\int\frac{d^3p}{(2\pi)^3}T_r[G(p,\mu)\gamma_4]\\
=&TN_cN_f\sum^{+\infty}_{-\infty}\int\frac{d^3p}{(2\pi)^3}\frac{4i\widetilde{\omega}_n}{p^2+M^2+\widetilde{\omega}^2_n}\\
=&\frac{TN_cN_f}{\pi^2}\int dp\cdot p^2\sum^{+\infty}_{-\infty}(\frac{1}{\epsilon-(i\omega_n-\mu)}-\frac{1}{\epsilon+(i\omega_n-\mu)})\\
=&\frac{N_cN_f}{\pi^2}\int dp\cdot p^2(\frac{1}{e^{\beta(\epsilon-\mu)}+1}-\frac{1}{e^{\beta(\epsilon+\mu)}+1}),
\end{split}
\end{eqnarray}
where $\epsilon=\sqrt{p^2+M^2}$. Then the quark number density of a single color and flavor is
\begin{equation}
n=\frac{1}{\pi^2}\int dp\cdot p^2(\frac{1}{e^{\beta(\epsilon-\mu)}+1}-\frac{1}{e^{\beta(\epsilon+\mu)}+1}).
\label{4}
\end{equation}
The variance of baryon number density is
\begin{eqnarray}
\begin{split}
\sigma^2 &=<(n_{B}-<n_{B}>)^2>\\
&=\frac{1}{3^2}<(n_{q}-<n_{q}>)^2>\\
&=\frac{1}{3^2}<[(n_u+n_d)-<n_u+n_d>]^2>\\
&=\frac{1}{3^2}<(n_u-<n_u>)^2>+<(n_d-n_d)^2>\\
&=\frac{N_f}{3^2}<(n_{f}-<n_{f}>)^2>.
\label{3}
\end{split}
\end{eqnarray}
where $n_f$ represents the quark number density of a single flavor. Since the up quarks and down quarks are independent, as in \cite{b9a}, we have $<(n_{u}-<n_{u}>)(n_{d}-<n_{d}>)>=0$ here. And $n_f=n_1+n_2+n_3$, where $n_1$, $n_2$ and $n_3$ represent the quark number density of three different color respectively. Since the confinement nature of QCD, $<n_in_j>=<n^2>$ ($i,j=1,2,3$). Then the variance of $n_f$ is
\begin{eqnarray}
\begin{split}
<(n_{f}-<n_{f}>)^2>&=<[(n_1+n_2+n_3)-<n_1+n_2+n_3>]^2>\\
&=N_c^2<(n-<n>)^2>,
\end{split}
\end{eqnarray}
and $<(n-<n>)^2>$ could be calculate in theory by the relation $<(n-<n>)^2>=T\frac{\partial n}{\partial \mu}=T\cdot\chi^{(2)}$. Then with Eq.{\ref{3}}, we could get
\begin{eqnarray}
\begin{split}
\sigma^2 &=\frac{N_f}{3^2}<(n_{f}-<n_{f}>)^2>\\
&=\frac{N_fN_c^2}{3^2}<(n-<n>)^2>\\
&=N_f<(n-<n>)^2>=N_f\cdot\chi^{(2)}.
\end{split}
\end{eqnarray}
Similarly, we could get
\begin{eqnarray}
\begin{split}
S\sigma &=\frac{<(n_{B}-<n_{B}>)^3>}{<(n_{B}-<n_{B}>)^2>}=\frac{N_f<(n-<n>)^3>}{N_f<(n-<n>)^2>}\\
&=\frac{T^2\chi^{(3)}}{T\chi^{(2)}}=T\frac{\chi^{(3)}}{\chi^{(2)}},
\end{split}
\end{eqnarray}
and
\begin{eqnarray}
\begin{split}
\kappa\sigma^2 &=\frac{<(n_{B}-<n_{B}>)^4>-3<(n_{B}-<n_{B}>)^3>}{<(n_{B}-<n_{B}>)^2>}\\
&=\frac{<(n-<n>)^4>-3<(n-<n>)^3>}{<(n-<n>)^2>}\\
&=T^2\frac{\chi^{(4)}}{\chi^{(2)}}.
\end{split}
\end{eqnarray}
where $\frac{\partial^2 n}{\partial \mu^2}=\chi^{(3)}$ and $\frac{\partial^3 n}{\partial \mu^3}=\chi^{(4)}$. That is to say by calculating the high order derivatives of Eq.{\ref{4}} with respect to $\mu$, we could get the experimental observables $S\sigma$ and $\kappa\sigma^2$ in the relativistic heavy ion collisions.

\section{Results}
\label{three}
Before comparing the $S\sigma$ and $\kappa\sigma^2$ results of the NJL model with the experimental data in the relativistic heavy ion collisions, we should find the correspondence of the freeze-out temperature $T$ and the quark chemical potential $\mu$ in the NJL results to the collision energies $\sqrt{S_{NN}}$ in the relativistic heavy ion collisions.

\begin{figure}
\centering
\includegraphics[width=1\linewidth]{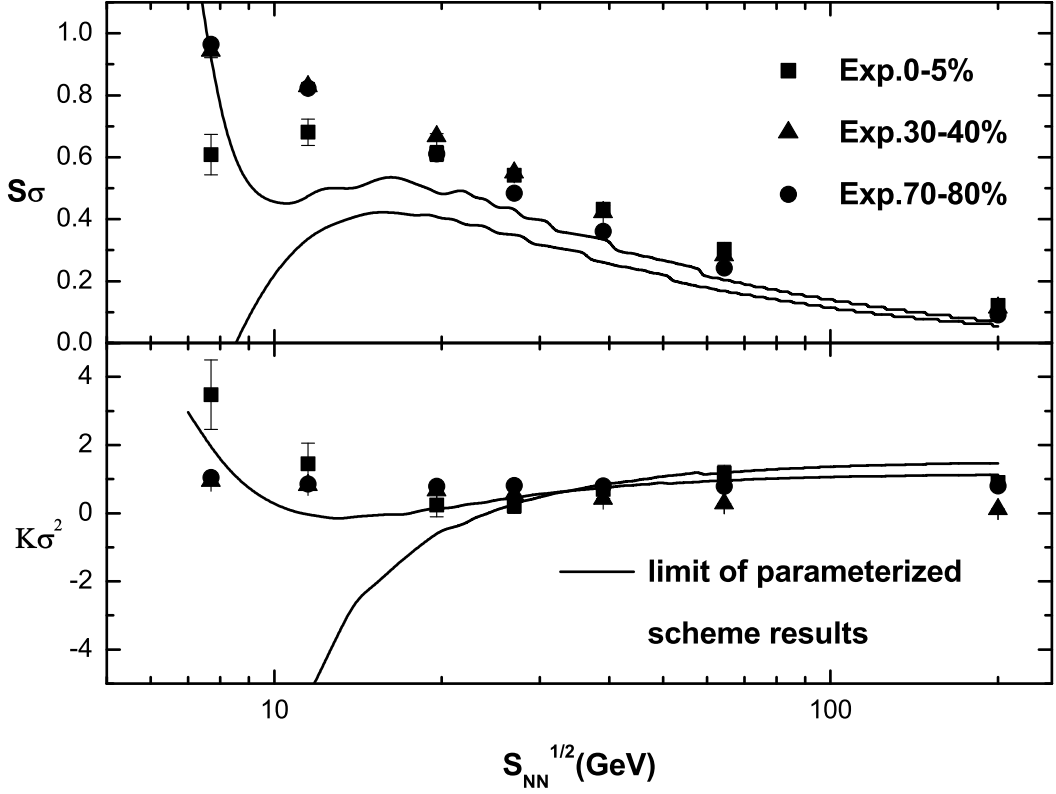}
\caption{Solid lines represent the upper and the lower limit of the NJL model results. Data points\cite{b11,b11a} are the experimental results of $S\sigma$ and $\kappa\sigma^2$($0-5\%$, $30-40\%$ and $60-80\%$ centralities).}
\label{njl}
\end{figure}

Firstly, we adopt a parameterized scheme by J. Cleymans and H. Osechler, etc in Ref.\cite{b10}. In the paper they proposed that $T=a-b\mu_B^2-c\mu_B^4$ and $\mu_B=d/(1+e\sqrt{S_{NN}})$, where $a=0.166\pm0.002 GeV$,$b=0.139\pm0.016 GeV^{-1}$,$c=0.053\pm0.021 GeV^{-3}$,$d=1.308\pm0.028 GeV$, $e=0.273\pm0.008 GeV^{-1}$ and $\mu_B=3\mu$ is the baryon chemical potential. Since each parameter(a,b,c,d,e) has the value range, we demonstrate the upper and the lower limit of the NJL model results in Fig.\ref{njl} as solid lines. At the same time, the $S\sigma$ and $\kappa\sigma^2$ for net-proton distributions\cite{b11,b11a} of three different centralities($0-5\%$, $30-40\%$ and $70-80\%$) at RHIC are shown as filled squares, triangles and circles respectively.

It is shown that the upper and the lower limit of the NJL model in Fig.\ref{njl} demonstrate an obvious fluctuation when the collision energies $\sqrt{S_{NN}}$ is less than $10 GeV$. And it is reduced when $\sqrt{S_{NN}}$ is more than $20 GeV$. That is to say, the NJL results is sensitive to the parameter of $T$ and $\mu$ when $\sqrt{S_{NN}}$ is less than $10 GeV$. Meanwhile, the NJL results are comprehensively less than the experimental data of both $S\sigma$ and $\kappa\sigma^2$, except the data of $\kappa\sigma^2$ when $\sqrt{S_{NN}}$ is more than $30 GeV$.
\begin{figure}
\centering
\includegraphics[width=1.0\linewidth]{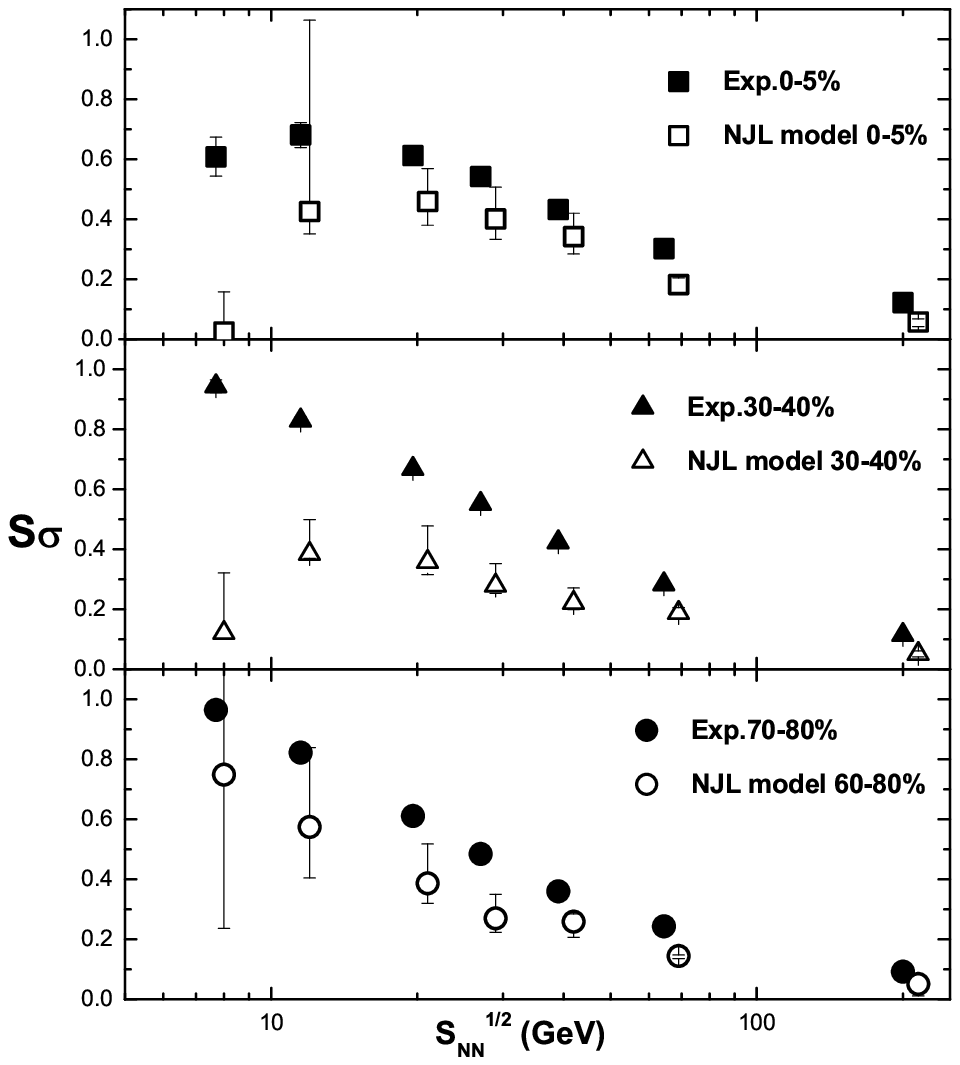}
\caption{$S\sigma$ of the NJL model and the experiments. The NJL results demonstrate as empty squares, triangles and circles($0-5\%$, $30-40\%$ and $60-80\%$ centralities respectively). And experimental data in Ref.\cite{b11,b11a} are shown as the filled patterns($0-5\%$, $30-40\%$ and $70-80\%$ centralities).}
\label{m1}
\end{figure}

\begin{figure}
\centering
\includegraphics[width=1.0\linewidth]{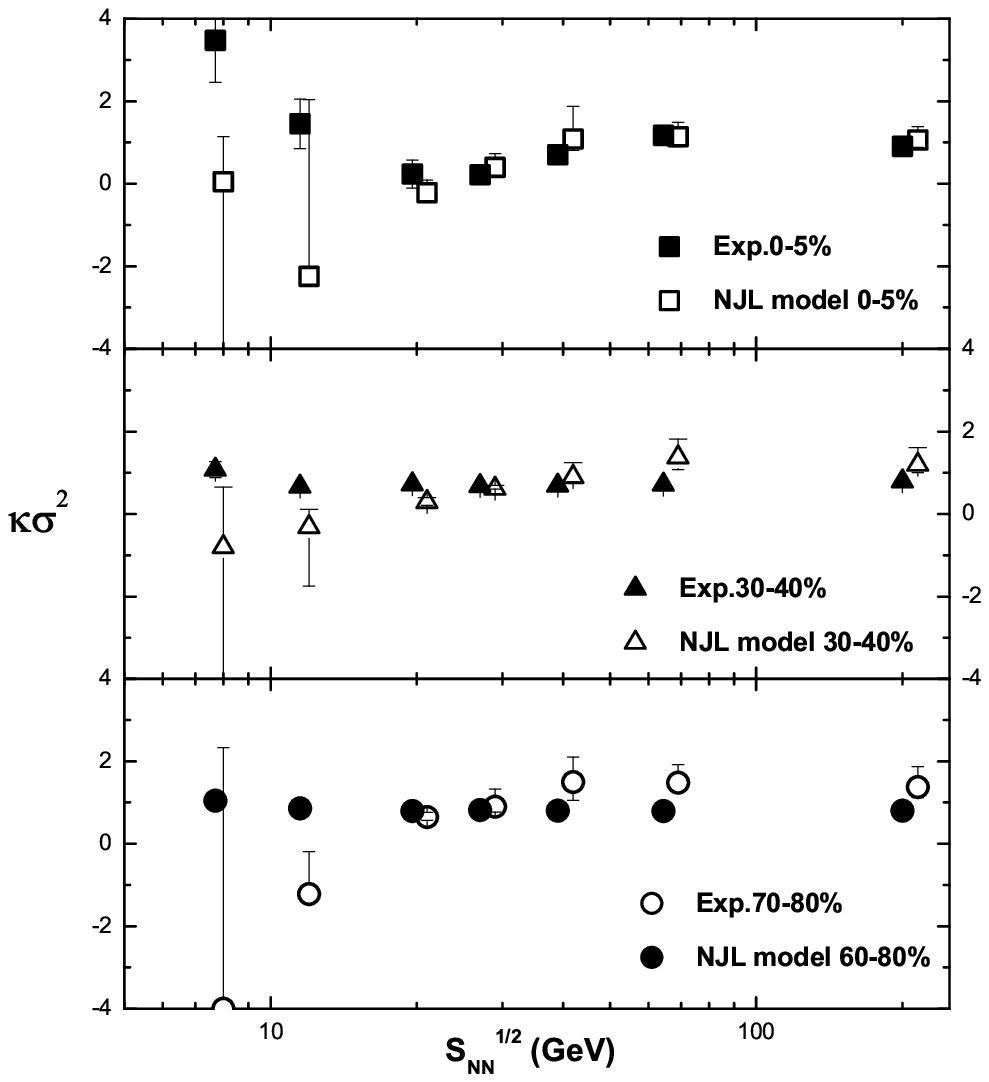}
\caption{$\kappa\sigma^2$ of the NJL model and the experiments. The NJL results demonstrate as empty squares, triangles and circles($0-5\%$, $30-40\%$ and $60-80\%$ centralities respectively). And experimental data in Ref.\cite{b11} are shown as the filled patterns($0-5\%$, $30-40\%$ and $70-80\%$ centralities).}
\label{m2}
\end{figure}

Secondly, the correspondence of the freeze-out $T$ and $\mu$ in the NJL results to the collision energies $\sqrt{S_{NN}}$ comes from the experimental results\cite{b12}. Since, in Ref.\cite{b12}, the correspondence is shown in three different centralities($0-5\%$, $30-40\%$ and $60-80\%$), then the NJL results of $S\sigma$ and $\kappa\sigma^2$ are also calculated in the three centralities and demonstrate as empty squares, triangles and circles in Fig.\ref{m1} and Fig.\ref{m2} respectively. And as in Fig.\ref{njl}, the experimental results in Ref.\cite{b11,b11a} are shown as filled patterns. It could be seen that the NJL result fit the experimental data better at large $\sqrt{S_{NN}}$ in Fig.\ref{m1} and Fig.\ref{m2}. And just as in Fig.\ref{njl}, the NJL results fluctuate obviously when the collision energies $\sqrt{S_{NN}}$ is less than $20 GeV$.

Now it might be concluded that the $S\sigma$ and $\kappa\sigma^2$ results of two-flavor NJL model fit the experiments better at large collision energies and show an obvious fluctuation at small collision energies. And this fluctuation reflect that the NJL model is sensitive to the parameters of $T$ and $\mu$ at small collision energies.

\section{Summary}
\label{four}

The moments of the distributions of conserved quantities, for example net-baryon number, in the relativistic heavy ion collisions are related to the correlation length $\xi$ of the system\cite{b3}, they are believed to be the good signatures of QCD phase transition and its CP. In order to cancel the volume term, the products of the moments, $S\sigma$ and $\kappa\sigma^2$, are constructed as the experimental observables. Since the moments of the baryon number are proportional to the various order baryon-number susceptibilities and  the quark-number density is determined by the corresponding dressed quark propagator only\cite{b6}, then by a reasonable dressed quark propagator we can calculate the moments of quark number at finite temperature and chemical potential.

We obtain the dressed quark propagator from the NJL model\cite{b7}, which is widely adopted to study the dynamical chiral symmetry breaking(DCSB) and the interaction between hadrons. When corresponding the freeze-out temperature $T$ and the quark chemical potential $\mu$ in the NJL results to the collision energies $\sqrt{S_{NN}}$ in the experiments, we choose the parameterized scheme by J. Cleymans and H. Osechler, etc and the RHIC results\cite{b12} at the same time. It is found that the $S\sigma$ and $\kappa\sigma^2$ results of two-flavor NJL model fit the experiments better at large collision energies and show an obvious fluctuation at small collision energies. And this fluctuation reflect that the NJL model is sensitive to the parameters of $T$ and $\mu$ at small collision energies.

\section*{Acknowledgements}

This work is supported by the University Natural Science Foundation of JiangSu Province China (under Grants No.17KJB140003 ).

\begin{appendix}
\section{}
The quark-number density susceptibilities($\chi^{(2)}$, $\chi^{(3)}$ and $\chi^{(4)}$) of a single color and flavor are:
\begin{equation}
\chi^{(2)}=\frac{\partial n}{\partial \mu}=-\frac{1}{\pi^2}\int dp\cdot p^2(\frac{e^s\frac{\partial s}{\partial \mu}}{(e^s+1)^2}-\frac{e^{s'}\frac{\partial s'}{\partial \mu}}{(e^{s'}+1)^2}),
\end{equation}

\begin{eqnarray}
\begin{split}
\chi^{(3)}=\frac{\partial^2 n}{\partial \mu^2}=&-\frac{1}{\pi^2}\int dp\cdot p^2(\frac{e^s(\frac{\partial^2 s}{\partial \mu^2}+(\frac{\partial s}{\partial \mu})^2)+e^{2s}(\frac{\partial^2 s}{\partial \mu^2}-(\frac{\partial s}{\partial \mu})^2)}{(e^s+1)^3}\\
&-\frac{e^{s'}(\frac{\partial^2 s'}{\partial \mu^2}+(\frac{\partial s'}{\partial \mu})^2)+e^{2s'}(\frac{\partial^2 s'}{\partial \mu^2}-(\frac{\partial s'}{\partial \mu})^2)}{(e^{s'}+1)^3}),
\end{split}
\end{eqnarray}

\begin{eqnarray}
\begin{split}
\chi^{(4)}=\frac{\partial^3 n}{\partial \mu^3}=&-\frac{1}{\pi^2}\int dp\cdot p^2(\frac{1}{(e^s+1)^4}(e^s(\frac{\partial^3 s}{\partial \mu^3}+3\frac{\partial s}{\partial \mu}\frac{\partial^2 s}{\partial \mu^2}+(\frac{\partial s}{\partial \mu})^3)\\
&+e^{2s}(2\frac{\partial^3 s}{\partial \mu^3}-4(\frac{\partial s}{\partial \mu})^3)+e^{3s}(\frac{\partial^3 s}{\partial \mu^3}-3\frac{\partial s}{\partial \mu}\frac{\partial^2 s}{\partial \mu^2}+(\frac{\partial s}{\partial \mu})^3))\\
&-\frac{1}{(e^{s'}+1)^4}(e^{s'}(\frac{\partial^3 s'}{\partial \mu^3}+3\frac{\partial s'}{\partial \mu}\frac{\partial^2 s'}{\partial \mu^2}+(\frac{\partial s'}{\partial \mu})^3)\\
&+e^{2s'}(2\frac{\partial^3 s'}{\partial \mu^3}-4(\frac{\partial s'}{\partial \mu})^3)+e^{3s'}(\frac{\partial^3 s'}{\partial \mu^3}-3\frac{\partial s'}{\partial \mu}\frac{\partial^2 s'}{\partial \mu^2}+(\frac{\partial s'}{\partial \mu})^3))),
\end{split}
\end{eqnarray}

where $s=(\epsilon-\mu)/T$ and $s'=(\epsilon+\mu)/T$, that is to say
\begin{eqnarray}
\begin{split}
\frac{\partial s}{\partial \mu}=\frac{1}{T}(\frac{\partial \epsilon}{\partial \mu}-1)=\frac{1}{T}(\frac{M}{\epsilon}\frac{\partial M}{\partial \mu}-1)\\
\frac{\partial s'}{\partial \mu}=\frac{1}{T}(\frac{\partial \epsilon}{\partial \mu}+1)=\frac{1}{T}(\frac{M}{\epsilon}\frac{\partial M}{\partial \mu}+1),
\end{split}
\end{eqnarray}

\begin{equation}
\frac{\partial^2 s}{\partial \mu^2}=\frac{\partial^2 s'}{\partial \mu^2}=\frac{1}{T}\frac{\partial^2 \epsilon}{\partial \mu^2}=\frac{1}{T}(\frac{1}{\epsilon}((\frac{\partial M}{\partial \mu})^2+M\frac{\partial^2 M}{\partial \mu^2})-\frac{1}{\epsilon^3}(M\frac{\partial M}{\partial \mu})^2),
\end{equation}

\begin{eqnarray}
\begin{split}
\frac{\partial^3 s}{\partial \mu^3}&=\frac{\partial^3 s'}{\partial \mu^3}=\frac{1}{T}\frac{\partial^3 \epsilon}{\partial \mu^3}\\
&=\frac{1}{T}(\frac{1}{\epsilon}(3\frac{\partial^2 M}{\partial \mu^2}\frac{\partial M}{\partial \mu}+M\frac{\partial^3 M}{\partial \mu^3})-\frac{3M}{\epsilon^3}\frac{\partial M}{\partial \mu}((\frac{\partial M}{\partial \mu})^2+M\frac{\partial^2 M}{\partial \mu^2})+\frac{3}{\epsilon^5}(M\frac{\partial M}{\partial \mu})^3),
\end{split}
\end{eqnarray}
and the value of $\frac{\partial M}{\partial \mu}$, $\frac{\partial^2 M}{\partial \mu^2}$ and $\frac{\partial^3 M}{\partial \mu^3}$ are obtained from the iterative equations
\begin{equation}
\frac{\partial M}{\partial \mu}=-2G_1\frac{\partial <\bar{\psi}\psi>}{\partial \mu}-4G_2<\bar{\psi}\psi>\frac{\partial <\bar{\psi}\psi>}{\partial \mu},
\end{equation}
\begin{equation}
\frac{\partial^2 M}{\partial \mu^2}=-2G_1\frac{\partial^2 <\bar{\psi}\psi>}{\partial \mu^2}-4G_2<\bar{\psi}\psi>\frac{\partial^2 <\bar{\psi}\psi>}{\partial \mu^2}-4G_2(\frac{\partial <\bar{\psi}\psi>}{\partial \mu})^2,
\end{equation}
\begin{equation}
\frac{\partial^3 M}{\partial \mu^3}=-2G_1\frac{\partial^3 <\bar{\psi}\psi>}{\partial \mu^3}-4G_2<\bar{\psi}\psi>\frac{\partial^3 <\bar{\psi}\psi>}{\partial \mu^3}-12G_2\frac{\partial <\bar{\psi}\psi>}{\partial \mu}\frac{\partial^2 <\bar{\psi}\psi>}{\partial \mu^2}.
\end{equation}

\end{appendix}

\end{document}